\documentclass[twoside,12pt]{article}
\usepackage{epsfig}

\newcommand{\be}{\begin{equation}}
\newcommand{\ee}{\end{equation}}
\newcommand{\bea}{\begin{eqnarray}}
\newcommand{\eea}{\end{eqnarray}}

\begin{document}

\title{ \vspace{1cm} The Spin of the Proton}
\author{A.~W.~Thomas$^{1,2}$ \\
$^1$Jefferson Lab, 12000 Jefferson Ave., Newport News VA 23606 USA \\
$^2$ College of William and Mary, Williamsburg VA 23187 USA \\}
\maketitle
\begin{abstract} 
The twenty years since the announcement of the proton spin crisis 
by the European Muon Collaboration has seen tremendous progress in 
our knowledge of the distribution of spin within the proton.  
The problem is reviewed, beginning with the original data and the 
suggestion that polarized gluons may play a crucial role in 
resolving the problem through the U(1) axial anomaly. The discussion 
continues to the present day where not only have 
strong limits have been placed on the amount of polarized glue in the 
proton but the experimental determination of the spin content has 
become much more precise. It is now clear that the origin of the discrepancy 
between experiment and the naive expectation of the fraction of spin 
carried by the quarks and anti-quarks in the proton lies in the 
non-perturabtive structure of the proton. We explain how the features 
expected in a modern, relativistic and chirally symmetric description 
of nucleon structure naturally explain the current data.
\end{abstract}
%\eject
%\tableofcontents

\section{Introduction}
In our quest to understand the structure of the 
nucleon and eventually of nuclei at the fundamental 
level of quarks and gluons, few problems have proven 
more exciting than that of mapping the distribution 
of the spin of the proton onto these 
constituents~\cite{Thomas:2001kw,Bass:2007zz}.
This began in earnest with the announcement by 
the European Muon Collaboration 
(EMC)~\cite{Ashman:1987hv} in 1988 that the 
quarks carried only a very small fraction of the 
proton spin, possibly zero. 
This soon became known as the "proton spin crisis". 

Several effects which 
were based on well known features of hadron structure 
were explored~\cite{Schreiber:1988uw,Myhrer:1988ap} 
but none were able to produce such a 
small number. On the other hand, it was soon 
realized that, because of the famous $U(1)$ 
axial anomaly~\cite{Adler:1969gk,Crewther:1978zz}, 
polarized gluons could also make a 
contribution to the proton spin 
structure 
function~\cite{Efremov:1988zh,Altarelli:1988nr,Carlitz:1988ab}. 
Indeed, it was shown that {\it if} the gluons 
in a polarized proton were to carry 4 units of 
spin, the EMC spin crisis would be resolved. No 
believable explanation of how the gluons might 
carry 4 units of spin was ever given. 
Nevertheless, the mathematical 
beauty of the proposed resolution   
was seductive and major experimental programs were 
launched to investigate it.

Over the past twenty years that experimental effort 
has produced spectacular results, with remarkable 
increases in precision and new kinds of data 
which probe for the presence of polarized gluons 
directly. There have also been some remarkable 
advances in our ability to solve QCD 
on a lattice of points in space-time and 
extract new information about nucleon structure. 
Indeed, the progress made has reached the stage 
where it is possible to assert that the gross 
features of the proton 
spin problem are now understood. 
This does not mean that 
considerable experimental and theoretical effort 
will not be required to confirm the resolution. 
However, as we shall explain, well known and 
understood aspects of hadron structure, in 
combination with the totality of modern experimental 
evidence leave little room for doubt.

The structure of this lecture is that we first 
review the experimental situation for the proton 
spin structure function as well as the search for 
polarized glue -- all in the context of the 
explanation proposed in terms of the axial 
anomaly. We then review the possible 
explanations in terms of more conventional 
aspects of hadron structure which were 
suggested within a few months of the EMC preprint. 
These explanations are then examined in the light 
of modern insights into hadron structure coming 
from the study of their properties as a function 
of quark mass using lattice QCD. We then close 
by summarizing the resolution of the spin problem 
and the future experimental work needed to 
confirm it.  

\section{The Experimental Data}
The major advance made by the EMC 
was to extend the 
measurements of the proton spin 
structure function,
$g_1^p(x,Q^2)$, to a value of Bjorken-$x$ 
almost as low as $10^{-2}$. This enabled a 
more accurate evaluation of the integral of the 
spin structure function. The latter can be 
rigorously expressed in terms of well known, 
perturbative QCD corrections and three low 
energy constants, the axial charges $g_A^{(3)}$ 
and $g_A^{(8)}$, which are known from 
neutron and hyperon 
$\beta$-decay, as well as the quantity $\Sigma$ 
(often written as $\Delta \Sigma$ or $a_0$)
\footnote{Here we denote by $\Sigma$ the renormalization 
group invariant spin content, $\Sigma(\mu^2=\infty)$. 
We refer to the discussion in Ref.~\cite{Myhrer:2007cf} of the
connection between $\Sigma$ calculated in a quark model 
and $\Sigma(\mu^2=\infty)$.} 
Within the parton model $\Sigma = \Delta u +
\Delta d + \Delta s$, with $\Delta q = \int_0^1 
\Delta q(x) dx$, the fraction of the helicity 
of the proton carried by the 
quarks and anti-quarks 
of flavor $q$. (We stress that 
while the word spin 
can be ambiguous, in the context 
of the spin problem 
it rigorously refers to the 
helicity of quarks in 
a proton, in the infinite momentum frame, with 
positive helicity.)

{}For clarity we simply denote the perturbative QCD 
coefficients, which are power series 
expansions in the strong coupling 
constant, $\alpha_s(Q^2)$,  
as $c_{i}(Q^2)$. Then 
the sum rule for the proton spin structure 
function is:
\be
\int_0^1 dx g_1^p(x,Q^2) = \frac{1}{12} 
c_1(Q^2) g_A^{(3)} + \frac{1}{36} c_1(Q^2) g_A^{(8)} 
+ \frac{1}{9} c_2(Q^2) \Sigma \, .
\label{eq:p-spin_sum}
\ee
Since $g_A^{(3,8)}$ are well known, having an 
accurate experimental determination of the 
left-hand side allows one to deduce $\Sigma$.
The result of the EMC measurement was 
\be
\Sigma = 14 \pm 3 \pm 10 \% \, ,
\label{eq:EMC_Sigma}
\ee
which is clearly incompatible with 100\%, or even 
65\%, which was the more realistic target after 
accounting for the relativistic motion of the 
quarks expected, for example, from bag model 
studies~\cite{Chodos:1974pn} -- see also Ref.~\cite{Cloet:2007em} 
for a similar result in a modern, relativistic confining 
model. Psychologically, it was 
even more important 
that this result was consistent with zero and 
this led to considerable excitement 
concerning the 
Skyrme model, which was initially believed to 
make just such a prediction. We note that it 
is now known that this is not the case and 
furthermore, as we shall see, the experimental data 
is no longer compatible with zero.

In modern terms Eq.~(\ref{eq:p-spin_sum}) is 
especially interesting as one can rigorously extract 
$\Sigma$ from a measurement of the axial 
weak, neutral 
current charge of the proton after correcting 
for heavy quark radiative 
effects~\cite{Bass:2002mv}. In this way 
it becomes a very interesting sum rule which can 
really only be violated if there is 
$\delta$-function contribution at $x=0$ -- as 
proposed by Bass~\cite{Bass:2004xa} 
in connection with the 
fascinating idea that some of the 
spin of a constituent quark might be associated with 
topological charge stored in its gluon fields.
At the time, the value of $\Sigma$ was immediately 
seen to violate the Ellis-Jaffe sum rule, which 
was based on the idea that if the proton contains 
no strange quarks one expects $\Sigma = g_A^{(8)}$; 
which is clearly violated by the EMC data. 
This led to the idea that 
the proton might have a significant polarized 
strange sea and that in turn stimulated a number of 
important experimental programs.

\subsection{The Possible Role of the U(1) Axial 
Anomaly}
Although the gluon carries no electric charge, 
it can create a quark-anti-quark pair which then 
interact with an incoming photon. If we view the 
deep-inelastic scattering cross section as the 
imaginary part of the forward photon-nucleon 
Compton amplitude, then this gluonic term corresponds 
to the imaginary part of a quark-anti-quark box 
diagram. The moments of the structure functions 
correspond to integrating over the momentum of 
the quark (or anti-quark) which is struck by the 
incoming photon and this effectively shrinks the 
box diagram to a triangle. In the case of spin 
dependent deep-inelastic scattering, 
the corresponding integral is quadratically 
divergent and the answer depends on 
whether or not the regularization procedure respects 
gauge invariance or chiral symmetry; it cannot 
respect both.

Thus, if one works within a scheme that ensures
chiral symmetry, the renormalized quark spin 
operators satisfy the usual SU(2) commutation 
relations ($[S_i,S_j] = \epsilon_{ijk} \, S_k$) 
and the axial current is conserved. 
On the other hand, 
if one chooses to ensure the more fundamental 
symmetry, namely gauge 
symmetry, the renormalized spin operators do not 
satisfy the usual commutation relations and the 
divergence of the renormalized axial current is 
not zero~\cite{Bass:2004xa,Bass:1992bk}. In that case, 
the value of $\Sigma$, 
defined through the matrix element of 
the axial charge, renormalized in a gauge invariant 
manner, can be written:
\be
\Sigma = \Sigma_{\rm naive} - 
\frac{N_f \, \alpha_s(Q^2)}{2 \pi} 
\Delta G (Q^2) \, .
\label{eq:DeltaG}
\ee
Here $\Sigma_{\rm naive}$ is the spin carried by the 
quarks in a naive quark model (i.e. one 
that respects chiral symmetry and ignores the 
effect of the axial anomaly) and the second term is 
the contribution arising, through the axial 
anomaly, from polarized gluons in the proton.

On the face of it, as $\alpha_s(Q^2)$ vanishes as 
$Q^2$ approaches $\infty$, it appears naively that 
the correction term on the right of 
Eq.(\ref{eq:DeltaG}) should vanish in the Bjorken 
limit. However, again because of the axial anomaly, 
the evolution of $\Delta G(Q^2)$ is such that the 
product, $\alpha_s(Q^2) \Delta G(Q^2)$, should 
go to a non-zero constant as 
$Q^2 \rightarrow \infty$. Thus the gluon box diagram 
must be included in the Bjorken limit.

At a typical scale of $Q^2=3$GeV$^2$, the 
strong coupling, $\alpha_s(Q^2)$, is around 0.3. 
Thus for 3 active flavors of quark the correction 
to the naive spin content is roughly $0.15 \, \Delta
G$ and {\it if} $\Delta G$ were of order 4, the 
naive expectation for the fraction of the spin of 
the proton carried by its quarks, namely about 2/3 
after allowance for the relativistic motion of the 
valence quarks, would be brought into agreement with 
the EMC data. 

It is important to realize a couple of things. 
First, no-one ever proposed a believable model 
of nucleon structure in which one could see how 
the gluons might carry so much spin. Secondly, 
it was understood that {\it if} the gluons carried 
4 units of spin they must also carry about 4 units 
of orbital angular momentum so that the total spin 
of the nucleon remains at 1/2. This second caution 
has tended to be forgotten as the limits on 
the amount of spin carried by gluons have tended 
to decrease.

Finally, we note that in the case of an explicitly 
gauge invariant scheme, such as $\overline{\rm MS}$,
there is no explicit gluon correction to 
$\Sigma$. Instead, the individual values of 
$\Delta q$ incorporate the effect of the axial 
anomaly. Whether one one works in a scheme that 
preserves chiral symmetry or gauge symmetry, 
provided the effect of the anomaly is treated 
consistently, the same physics is incorporated. 
It just appears in different places.   

\subsection{The Shape of the Gluon Contribution
to $g_1^p$}
Although most of the focus on the gluon contribution 
to the spin structure function of the proton focussed 
on its first moment, it is also of considerable 
importance to know the characteristic 
shape of 
the contribution from the photon-gluon box 
diagram to the proton spin 
structure function~\cite{Bass:1991yx}. In 
fact, as emphasized very early by Bass and 
co-workers, the shape of this contribution is 
distinctive, with $\delta g_1^p(x)$ going rapidly 
large and negative for $x$ 
below $10^{-2}$~\cite{Bass:1992bk}  -- if 
$\Delta G$ is positive, as originally 
required to explain the 
spin crisis.

We also note that $\Sigma$ is a flavor singlet 
quantity, so it will be the same for the proton 
and the neutron. Thus, while the large iso-vector  
contribution (corresponding to $g_A^{(3)}$ in 
Eq.(\ref{eq:p-spin_sum}) ) cancels in the 
measurement of the deuteron spin structure function, 
the contribution of the photon-gluon box diagram 
is enhanced. This makes the recent, very precise
measurements of the deuteron spin structure function 
down to below $10^{-4}$, by the COMPASS 
Collaboration~\cite{Ageev:2007du,Ageev:2005gh}, 
especially significant. Those
measurements are completely consistent with zero 
for $x$ from $5 \times 10^{-5}$ to almost $10^{-1}$. 
This alone puts a very severe constraint on the 
experimental value of $\Delta G$ -- making it 
impossible for $\Delta G$ to be anywhere near 
as large as 4. 

At Jefferson Lab, the CLAS Collaboration has 
more than doubled the number of polarized spin 
structure function data in the deep-inelastic region.
Because of the relatively low range of $Q^2$ it 
is crucial to allow for higher-twist effects in 
amalyzing the data. Nevertheless, Leader 
{\it et al.}~\cite{Leader:2006xc} have found 
that including the JLab 
data~\cite{Dharmawardane:2006zd} in their global 
analysis reduces the error on 
the determination of $x \Delta G$ by a factor of 
3 from what is possible without it. Their 
conclusion is that $|\Delta G| < 0.3$.
 
\subsection{Direct Searches for $\Delta G$}
Another characteristic of the gluon box diagram 
is that it tends to be dominated by quark-anti-quark 
pairs with transverse momentum equal to or larger 
than $Q$~\cite{Carlitz:1988ab}. Thus, one can 
search directly for 
the existence of polarized gluons by making 
semi-inclusive measurements that explicitly 
look for high-$p_T$ hadrons in a deep-inelastic 
event. 

Such direct measurements have been extensively 
explored at Hermes~\cite{Liebing:2007zz} and 
COMPASS~\cite{Ageev:2005pq,Kabuss:2006zx}, with negative 
results. For two different methods of analysis, 
Hermes has found $\Delta G / G$ to be of order 
$0.071 \pm 0.034 \pm 0.011$ and $0.071 \pm 0.034 
\pm 0.010$ in the low-$x$ region. This led them 
to the conclusion that $\Delta G$ is small 
and ``unlikely to solve the puzzle of the 
missing nucleon spin''. The COMPASS result is 
completely consistent with Hermes but even more 
precise, with the result $\Delta G / G = 0.06
\pm 0.31 \pm 0.06$ at $x = 0.13$ and 
$\mu^2 = 3$GeV$^2$, using data with 
$Q^2 > 1$GeV$^2$ and looking at high-$p_T$ hadrons.

At RHIC both PHENIX~\cite{PHENIX_1,Adare:2007dg} 
and STAR~\cite{STAR_1,Abelev:2007vt} have measured  
asymmetries in polarized proton-proton collisions, 
either looking for asymmetries (i.e. differences in 
the cross sections for helicities anti-parallel 
and parallel) in either jet or $\pi^0$ production. 
The preliminary analysis of the latest PHENIX 
data prefers a value of $\Delta G$ between -0.5 
and zero. For STAR the preliminary analysis of Run 6 
yielded a limit for $|\Delta G|$ below 0.3 (at 
90\% confidence level) and again 
consistent with zero. 

\subsection{Progress in Determining the Integral 
of $g_1^p(x,Q^2)$}
The tremendous progress in the measurement of 
both the proton and neutron spin structure functions 
over the twenty years since the original EMC 
discovery have led to a major increase in 
the accuracy with which the integral of $g_1^p$ 
can be determined. We have already mentioned 
that the range of $x$ over which data exists 
now extends below $10^{-4}$. In addition, in 
comparison with the Spin Muon Collaboration, the 
successor to EMC at CERN which was designed 
to follow up the discovery of the spin crisis, 
the latest data is as much as an order of 
magnitude more precise.
The latest analyses of Hermes~\cite{Airapetian:2007mh} 
and COMPASS~\cite{Alexakhin:2006vx} yield 
values:
\bea
\Sigma &=& 0.330 \pm 0.011 {\rm (thry)} \pm 0.025
{\rm (exp)} \pm 0.028 {\rm (evol)} \, \, 
\, \, {\rm Hermes} \nonumber \\
\Sigma &=& 0.33 \pm 0.03 {\rm (stat)}  \pm 
0.05 {\rm (syst)} \, \, \, \, \, \, \, \, \, \, \, 
\, \, \, \, \, \, \, {\rm COMPASS} \, .
\label{eq:data}
\eea
This represents a very substantial increase in the 
fraction of the spin of the proton carried by its 
quarks and anti-quarks. 

\subsection{Summary of the Experimental Situation}
The result of the last twenty years of intense 
experimental effort has been a very significant 
improvement in the accuracy with which $\Sigma$ 
is known. Contrary to the original EMC measurement 
it is no longer possible for the quarks to carry 
none of the spin of the proton. Rather it seems 
that the quarks and anti-quarks in the proton 
carry about a third of its spin and possibly as 
much as 40\%.

As well as this shift in the target which theory 
needs to explain, the experimental effort has also 
established very strong upper limits on the 
spin of the gluons. It seems that $\Delta G$ cannot 
be larger than 0.3 and may even be slightly negative.
This value of $\Delta G$ is more than an 
order of magnitude larger than the 
value (namely 4) originally 
found necessary to reduce the naive theoretical 
expectation of 65\% (after allowing for relativistic
motion of the valence quarks) to the value 
observed by EMC, through the axial anomaly. 
Through Eq.~(\ref{eq:DeltaG}) we see that 
$|\Delta G| < 0.3$ implies that through the axial 
anomaly the gluons yield a correction of 
less than 5\% to the quark spin content of the proton.
Thus the gluon spin is also a factor of 6 
too small to explain 
the difference between 65\% and the current 
experimental values of $\Sigma$, given in 
Eq.~(\ref{eq:data}). Indeed, if the hints from 
PHENIX are right and $\Delta G$ is negative, it 
may actually make the problem 
marginally worse, rather than 
helping to resolve it.

The nature of the challenge presented by the proton 
spin structure function has changed but it is 
still a fascinating problem.

\section{The Modern Theoretical Explanation}
The discussion in this section has its origins 
in two papers written in 1988. The first, written 
with Andreas Schreiber~\cite{Schreiber:1988uw}, 
dealt with the effect of the 
pion cloud of the nucleon, which is well known to 
be associated with chiral symmetry. The second, 
written with Fred Myhrer~\cite{Myhrer:1988ap}, 
dealt with what in nuclear 
physics would be described as an exchange current 
correction -- in this case arising through the 
hyperfine one-gluon-exchange (OGE) interaction 
which is usually considered primarily responsible 
for the difference in mass between the nucleon 
and $\Delta$, the $\Sigma$ and $\Lambda$ and so 
on. We consider these two terms in order, noting 
that the discussion in this section follows 
closely that presented recently by Myhrer and 
Thomas~\cite{Myhrer:2007cf}.

\subsection{The Role of the Pion Cloud}
That virtual pion emission and absorption can 
play a major role in the properties of hadrons 
has been appreciated since its existence was 
proposed by Yukawa. Within the modern context 
of QCD as the fundamental theory of the strong 
interaction this remains true, even though we 
know that the pion is itself composed of 
quark-anti-quark pairs. Indeed, we understand that 
the unusually small mass of the pion is associated 
with the very low values of the $u$ and $d$ quark 
masses, with $m_\pi^2$ vanishing linearly as the 
light quark masses go to zero. That is, it is a 
pseudo-Goldstone boson associated with the 
approximate chiral invariance of QCD.

This special property of the pion leads to many 
important consequences. For example, in the chiral 
limit (vanishing light quark masses) the 
charge radius of the proton and neutron go to 
infinity (plus and minus, respectively). 
Chiral perturbation theory provides a systematic 
way of exploring the consequences of chiral 
symmetry. In particular, it has established that 
hadron properties, such as charge radii, magnetic 
moments and so on, are non-analytic functions of 
the light quark masses solely because of the 
contributions of Goldstone boson loops. Tracing 
the non-analytic behaviour of hadron properties 
has proven a powerful tool in guiding the development 
of models and in testing whether 
apparent relationships between physical observables 
are purely accidental or whether they 
may be somewhat deeper~\cite{Leinweber:2001ui}.

An unfortunate consequence of the mathematical 
elegance of the  formalism of chiral perturbation 
theory is that it has been allowed to obscure a 
number of instances where there is a simple,  
physically meaningful explanation of some 
interesting physical phenomena. 
We recently presented arguments as to why it 
is both meaningful and even satisfying to think 
of hadron structure in terms of a pion cloud and 
we refer interested readers to 
that discussion~\cite{Thomas:2007bc}.
 
In fact, describing a physical nucleon as having 
a pion cloud which interacts with the valence 
quarks of the quark core (the  ``bare'' nucleon), 
in a manner dictated by the requirements of 
chiral symmetry, has been very successful 
in describing the properties of the 
nucleon~\cite{Theberge:1980ye,Thomas:1982kv}. 
The cloudy bag model
(CBM)~\cite{Theberge:1980ye,Thomas:1982kv} 
reflects this description of the nucleon and in 
this model the nucleon consists of a bare nucleon, 
$| N>$, with a probability 
$Z \sim 1-P_{N\pi} -P_{\Delta \pi} \, 
\sim \, 0.7$, in addition to being described 
as a nucleon ($N$) and a pion and a $\Delta$ 
and a pion, with probabilities 
$P_{N\pi} \sim 0.20-0.25 $ and 
$P_{\Delta \pi} \sim 0.05-0.10 $, respectively. 
The phenomenological constraints on these
probabilities were discussed, for example, 
in Refs.~\cite{Thomas:2007bc,Speth:1996pz,Melnitchouk:1998rv}. 
One of the most famous of these constraints 
is associated with the excess of $\bar{d}$ 
over $\bar{u}$ quarks in the proton, 
predicted on the basis of the CBM~\cite{Thomas:1983fh}. 
Indeed, to first 
order the integral of $\bar{d}(x) - \bar{u}(x)$ 
is $2/3 P_{N\pi} \, - P_{\Delta \pi}/3$, 
which is consistent with the experimental 
data~\cite{Arneodo:1996kd} if $P_{N\pi}$ and 
$P_{\Delta \pi}$ lie within the ranges just 
quoted.

The effect of the pion cloud on the spin carried 
by quarks was investigated very 
early by Schreiber and Thomas, who wrote 
the corrections to the spin sum-rules for the 
proton and neutron explicitly in terms of the 
probabilities set out above~\cite{Schreiber:1988uw}. 
For the present purpose it is helpful to rewrite 
the results of Ref.~\cite{Schreiber:1988uw} for 
the proton and neutron. In fact, if we consider 
explicitly the flavor singlet combination, 
proton plus neutron, the pion cloud modifies the 
quark contribution to the proton spin in 
the following manner:
\be
\Sigma \rightarrow  \left(Z - \frac{1}{3} 
P_{N \pi} +\frac{5}{3} P_{\Delta \pi} \right) 
\Sigma \, .
\label{eq:pion}
\ee

The critical feature of the role of 
the pion cloud shown in Eq.~(\ref{eq:pion}) is 
that the Clebsch-Gordon algebra for coupling 
the spin of the nucleon and the orbital angular 
momentum of the pion in the $N \pi$ Fock 
state favors a spin down nucleon and a pion 
with +1 unit of orbital angular momentum 
in the $z$-direction. 
This has the effect of replacing quark spin by 
quark and anti-quark orbital angular momentum. 
Note that in the $\Delta \pi$ Fock component 
the spin of the baryon tends to point up 
(and the pion angular momentum down), 
thus enhancing the quark spin. 
Nevertheless, the wave function renormalization 
factor, $Z$, dominates, yielding a reduction 
by a factor between 0.7 and 0.8 for the range 
of probabilities quoted above.

\subsection{The Role of the One-Gluon-Exchange Hyperfine Interaction}
It is well established that the spin-spin 
interaction between quarks in a baryon, arising from the 
exchange of a single gluon, explains a 
major part of the mass difference between the 
octet and decuplet baryons -- e.g., 
the N-$\Delta$ mass 
difference~\cite{Chodos:1974pn,De Rujula:1975ge}.
This spin-spin interaction must therefore also 
play a role when an external probe interacts with 
the three-quark baryon state. That is, the probe 
not only senses a single quark current but 
a two-quark current as well. 
The latter has an intermediate quark propagator connecting the probe and 
the spin-spin interaction vertices, 
and is similar to 
the exchange-current corrections which are 
well known in nuclear physics. 
\begin{figure}[h]
%\vspace{-0.5cm}
\centering
\includegraphics[width=8cm]{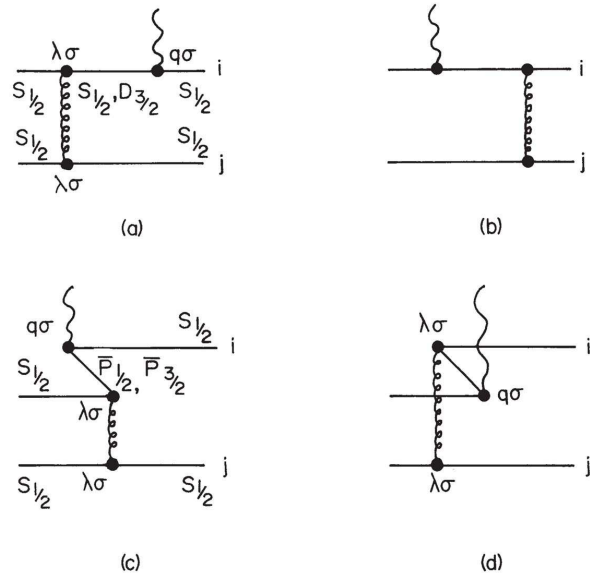}
\caption{The one-gluon-exchange correction to the 
spin sum rule investigated by 
Myhrer and Thomas~\protect\cite{Myhrer:1988ap,Myhrer:2007cf}.
}
\label{fig:OGE}
\end{figure}

In the case of spin dependent properties, 
the probe couples to the 
various currents in the nucleon. 
In the first exploration of the 
two-quark current, illustrated in Fig.~\ref{fig:OGE},  
carried out by Hogaasen and 
Myhrer~\cite{Hogaasen:1988jd}, the MIT bag model 
was used and the quark propagator was written as a 
sum over quark eigenmodes. The 
dominant contributions were found to come from 
the intermediate p-wave anti-quark states. 

The primary focus of Ref.~\cite{Hogaasen:1988jd} was 
actually the OGE corrections 
to the magnetic moments and semi-leptonic 
decays of the baryon octet. For example, this 
exchange current correction is vital to 
understand the unusual strength of 
the decay $\Sigma^- 
\rightarrow n + e^- + \bar{\nu}_e$. 
Myhrer and Thomas~\cite{Myhrer:1988ap} 
realized the importance of this correction to 
the flavor singlet axial charge and hence to the proton 
spin, finding that it reduced the fraction of the spin of 
the nucleon carried by quarks,  calculated in the naive 
bag model, by 0.15 -- i.e., 
$\Sigma \rightarrow \Sigma -3G $~\cite{Myhrer:1988ap}. 
The correction term, $G$, is proportional to $\alpha_s$ times 
certain bag model matrix elements~\cite{Hogaasen:1988jd}, 
where $\alpha_s$ is determined by 
the ``bare" nucleon-$\Delta$ mass difference. 

As in the case of the correction for relativistic 
motion of the valence quarks and for the pion 
cloud, here too the spin lost by the quarks is compensated 
by orbital angular momentum -- here it is orbital 
angular momentum of the quarks and anti-quarks 
(the latter predominantly $\bar{u}$ in the p-wave).

\subsection{Summary of the Theoretical Corrections}
We have examined three corrections to the fraction 
of the spin of a nucleon carried by its quark 
and anti-quarks that arise naturally in any realistic 
treatment of proton structure:
\begin{itemize}
\item The valence quarks must be treated 
relativistically. For example, in the MIT bag model 
they satisfy the Dirac equation and the lower 
component of the Dirac spinor for a spin-up valence 
quark in an s-state will have predominantly 
$L_z \, = \, +1$ and spin down. This reduces the 
fraction of the spin carried by valence quarks 
from 100\% to around 65\%.
\item The Clebsch-Gordon algebra for the dominant 
Fock component of the nucleon wave function 
associated with its pion cloud is identical. That is, 
the dominant term has the quark core (or ``bare 
nucleon'', with spin down) while the pion has 
predominantly $L_z \, = \, +1$. Within the context 
of chiral quarks models, such as the CBM, this means 
that whatever the spin content within the quark model 
used, dressing it with the pion cloud reduces that 
value by a factor of order (0.7,0.8).
\item The exchange current correction associated 
with OGE in the proton also reduces the amount 
of spin carried by the quarks by about 15\%, where 
this number has so far only been calculated 
within the MIT bag model.
\end{itemize}

Clearly, if we combine these three corrections, that 
is we assume a relativistic description of the  
structure of the nucleon which incorporates OGE 
and chiral symmetry, the simple accounting of 
the fraction of the nucleon spin carried by 
its quarks and anti-quarks is:
\bea
\Sigma &=& (0.7,0.8) \times (0.65 \, - \, 0.15)
\nonumber \\
&\rightarrow& \Sigma \in (0.35, 0.40) \, .
\label{eq:accounting}
\eea
This result is in very satisfactory agreement with 
the current experimental values given earlier in 
Eq.~(\ref{eq:data}) and the problem appears to be 
solved -- at least at the level of a few percent.

There were two reasons why this conclusion could not 
be drawn in 1988. First the experimental target was 
14\% and possibly zero and even combined these 
corrections could not explain that. Secondly, at the 
time, the various studies of chiral quark models 
suggested that a large fraction of the mass 
splitting between the N and $\Delta$ arose from 
pion cloud corrections. If that were so, it would 
be double counting to include both the pion cloud 
correction and the OGE correction with a strength 
determined by the observed N-$\Delta$ mass 
splitting.

Experimental progress over the past 20 years 
has eliminated the first problem. For the second 
we owe a resolution to the sophisticated studies 
of hadron properties as a function of quark mass 
that have been stimulated by lattice QCD over 
the past decade. We review this briefly in the next 
section, showing that as a result of discoveries 
by Leinweber, Young and Thomas concerning the 
link between quenched and full lattice QCD, we 
now know that only $40 \pm 20$ MeV of the 
observed 300 MeV N-$\Delta$ mass splitting comes 
from the pion cloud~\cite{Young:2002cj}. 
Consequently, there is 
little or no double counting as 80-90\% of the 
N-$\Delta$ mass difference would then come from 
OGE and the effective strong coupling constant,
$\alpha_s$, used to calculate the OGE correction 
to the quark spin is appropriate.

\section{The Origin of the N-$\Delta$ Mass 
Difference and Lattice QCD}
One of the unexpected but very positive 
consequences of our {\it lack} of 
supercomputing power is the fact that it has not 
been possible to compute physical hadron properties 
in lattice QCD. In fact, with computation time 
scaling like $m_\pi^{-9}$ (if we include the 
larger lattice size needed), calculations 
have covered the pion mass range from 0.3 
to 1.0 GeV (or higher). Far from being 
a disappointment, this has given us a wealth 
of unexpected insight into how QCD behaves 
as the light quark masses are 
varied~\cite{Thomas:2002sj}. In terms of 
the insight this has given us into hadron 
structure it is both truly invaluable and 
thus far under-utilized. 

The most striking feature of the lattice data 
is that in the region $m_\pi > 0.4$ GeV, 
in fact for almost all of the simulations 
made so far, all baryon properties show a 
smooth dependence on quark mass, 
totally consistent with that expected 
within a constituent quark 
model. The rapid, non-linear dependence 
on $m_\pi$ required by the LNA and NLNA 
behavior of $\chi$PT are notably absent from the 
data!

The conventional view of $\chi$PT has no 
explanation for this simple, universal 
observation. Worse, in seeking to apply 
$\chi$PT to extrapolate the data back to 
the physical pion mass, it has been necessary 
to rely on ad hoc cancellations between 
the high order terms in the usual power 
series expansion (supplemented by the 
required non-analytic behavior). In fact, 
there is strong evidence that such series 
expansions have been applied well beyond 
their region of 
convergence~\cite{Leinweber:2005xz} 
and that as a result such extrapolations are 
largely unreliable.
\begin{figure}[t]
\vspace{-0.5cm}
\centering
\includegraphics[width=17.0cm]{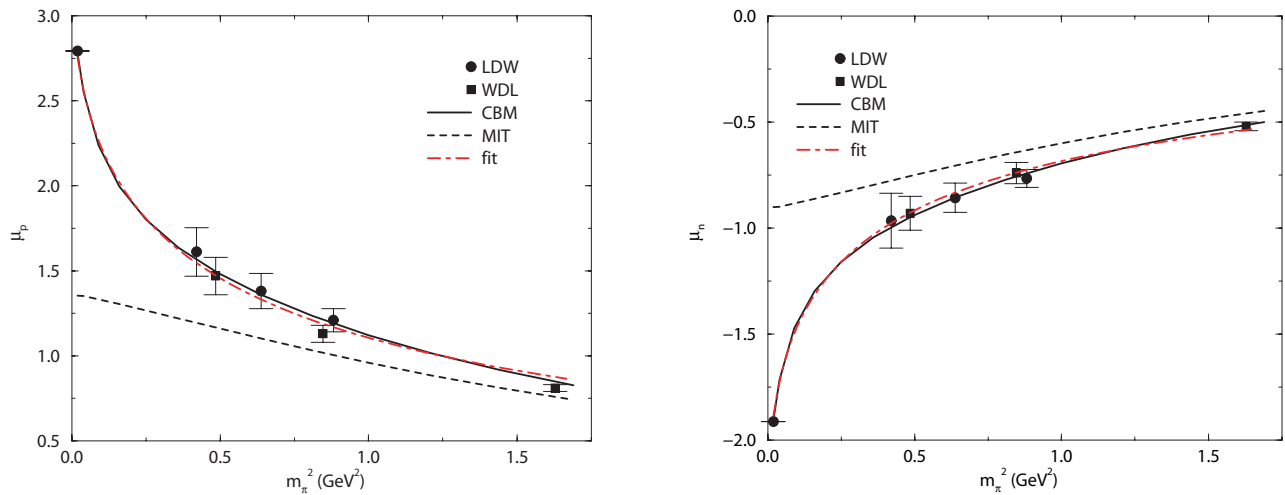}
\caption{Comparison between early lattice QCD  
data for $\mu_p$ and $\mu_n$ 
and the CBM versus 
$m_\pi^2$~\protect\cite{Leinweber:1998ej}. 
The model is also compared with a simple chiral fit 
and the MIT bag model (without the pion cloud).}
\label{MagMom}
\end{figure}

On the other hand, the picture of the pion cloud 
that we have discussed briefly yields a very 
natural explanation of the universal, 
constituent quark model behaviour 
of hadron properties found in the lattice 
simulations for $m_\pi > 0.4$ GeV. 
The natural high momentum cut-off on the 
momentum of the emitted pion, 
which is associated with the finite size 
(typically $R \sim$ 1 fm) of the bare baryon 
(i.e., the bag in the CBM), strongly suppresses 
pion loop contributions as $m_\pi$ increases. 
The natural mass scale which sets the 
boundary between rapid chiral variation 
and constituent quark type behavior is 
$1/R \sim 0.2 \, {\rm to} \, 0.4$ GeV. 
Indeed, when in the early investigation of the 
quark mass dependence of nucleon properties 
the CBM was compared directly with lattice 
data, the agreement was remarkably good~\cite{Leinweber:1999ig}. 
(Similar results have been obtained recently within 
the chiral quark soliton model~\cite{Goeke:2005fs}.) 
We illustrate this in Fig.~\ref{MagMom} for 
the magnetic moments of the proton and neutron. 
The results were equally as impressive for 
the N and $\Delta$ masses and 
magnetic moments, the proton charge radius 
and the moments of its parton distribution 
functions~\cite{Detmold:2002nf}. The 
key features necessary to reproduce 
the behaviour found at large quark mass 
in lattice QCD {\em and} to reproduce 
the experimentally measured data at 
the physical mass 
seem to be that:
\begin{itemize}
\item The treatment of the pion cloud (chiral) 
corrections ensures the correct LNA (
and NLNA, although in practice this seems less 
important in many applications) behaviour of QCD 
\item The pion cloud contribution is 
suppressed for $m_\pi$ beyond 0.4 GeV, and 
\item the underlying quark model exhibits 
constituent quark like behaviour for the 
corresponding range of current quark masses.
\end{itemize}

While the CBM satisfies all of these properties,     
in analyzing lattice 
QCD data one does not want to rely on any 
particular quark model. However, one does 
need to suppress the pion cloud as 
$m_\pi$ goes up, and the simple use of 
a finite range regulator (FRR) in the 
evaluation of the pion loops that yield 
the LNA and NLNA behaviour ensures 
this at the cost of one additional parameter, 
the cut-off mass $\Lambda$. If the data 
are good enough one can use this as 
a fitting parameter but in general 
it is sufficient~\cite{Lepage:1997cs} to 
choose a value consistent with the physical 
arguments presented above. 
The sensitivity of the extrapolation to 
the choice of the functional form of 
the FRR is then an additional source 
of systematic error in the final quoted result. 
In the case of the nucleon mass the corresponding 
systematic error was~\cite{Leinweber:2003dg} 
of the order of a mere 0.1\%.

One of the most remarkable results of this 
physical understanding of the role 
of the pion cloud and, in particular, 
its suppression at large pion mass has been 
the unexpected discovery of a connection 
between lattice simulations 
based upon quenched QCD (QQCD) 
and full QCD~\cite{Aoki:1999yr}. 
In a study of the quark mass dependence of 
the N and $\Delta$ masses~\cite{Young:2002cj}, 
it was discovered that if the 
self-energies appropriate to either QQCD 
or full QCD were regulated using the same 
dipole form for the FRR (the dipole being 
the most natural physical choice given that 
the axial form factor of the nucleon has 
a dipole form) with mass parameter 
$\Lambda = 0.8$ GeV (the preferred value, 
as noted above), then the residual expansions for 
the nucleon mass in QQCD and QCD (and also for 
the $\Delta$ in QQCD and QCD) were the same 
within the errors of the fit! This is a 
remarkable result which a posteriori gives 
enormous support to the physical picture of 
the baryons consisting of confined valence 
quarks surrounded by a perturbative pion cloud. 
The baryon core is basically determined by 
the confinement mechanism and provided the choice 
of lattice scale reproduces the physically 
known confining force (either through the 
string tension or the 
Sommer parameter~\cite{Sommer:1993ce}, 
derived from the heavy quark potential) 
it makes little difference whether one uses 
QQCD or full QCD to describe that core. 
What {\em does} matter is the change 
in the chiral coefficients as one goes from 
QQCD to full QCD.

Perhaps the most significant application of 
this discovery has been the application 
to the calculation of the octet magnetic 
moments and charge radii based on 
accurate QQCD simulations that extend to 
rather low quark mass. Using the constraints 
of charge symmetry this has led to some extremely 
accurate calculations of the strange quark 
contributions to the magnetic 
moment~\cite{Leinweber:2004tc} and charge 
radius~\cite{Leinweber:2006ug} of the 
proton. Indeed, those calculations 
are in excellent agreement with 
the current world 
data~\cite{Young:2006jc,Young:2007zs} 
but, in a unique example in modern strong 
interaction physics, they are an order 
of magnitude more accurate.

In the present context, the key result of the 
analysis of QQCD and full QCD data for the 
N and $\Delta$ masses is that the contribution of 
the pion cloud to their mass difference is 
really under control. The result is that only 
$40 \pm 20$ MeV of the observed mass difference 
can be attributed  to pions. The rest must 
be associated with other mechanisms such as the 
traditional OGE hypefine interaction.

\section{Summary and Discussion}  
Two decades of experimental investigation have 
given us a wealth of important new information 
concerning the spin structure of the proton. We 
now know that the spin crisis is nowhere near 
as severe as once thought but still only about 
a third of the proton spin is carried by its quarks.
Polarized gluons, which in principle could contribute 
through the axial anomaly, in practice seem to play 
no significant role. It seems likely that less 
than 5\% of the missing spin can come from 
polarized glue and the sign may be such that it 
makes the problem slightly worse.  

Instead it appears that important aspects of the 
non-perturbative structure of the nucleon {\it do} 
resolve the crisis. Indeed, three pieces of 
physics present in any realistic model are required. 
These consist of the relativistic motion of the 
valence quarks, the one-gluon-exchange  
interaction needed to describe the hyperfine 
splitting of hadron masses (especially the 
N-$\Delta$ mass difference) and finally the 
inclusion of the pion cloud required by chiral 
symmetry. These three terms reduce the fraction 
of the spin of the proton carried by its quarks to 
between 35 and 40\%, in excellent agreement with 
the latest experimental data.
The theoretical consistency of this picture owes 
a great deal to recent studies of the dependence 
on quark mass of hadron properties calculated in 
lattice QCD. 

For the future, it will be critical to test that 
the missing components of the proton spin do 
indeed reside as orbital angular momentum on the 
quarks and anti-quarks, as implied by this 
theoretical explanation. In this respect the 
program of measurements of Generalized 
Parton Distributions, especially following 
the 12 GeV Upgrade at Jefferson Lab, will be 
vital~\cite{12GeV,Guidal:2004nd,Belitsky:2005qn}.

\section*{Acknowledgements}
I would like to thank both Andreas Schreiber and 
Fred Myhrer who were partners in the original 
work on the role of non-perturbative structure in 
the spin crisis. The collaboration with Steven Bass has been 
important to my understanding of the role of 
gluons in the spin problem. Finally, the insight 
into the behaviour of hadron properties as a 
function of quark mass owes a great deal to Derek 
Leinweber and Ross Young. This work was supported 
in part by U.~S.~DOE Contract No. DE-AC05-06OR23177, 
under which Jefferson Science Associates, LLC, 
operates Jefferson Lab.

\end{document}